\documentclass[aps,prl,twocolumn,superscriptaddress]{revtex4}

\usepackage{amsmath}
\usepackage{graphicx}
\usepackage{bm}

\begin{document}

\title{Unexpectedly robust protection from backscattering in the
topological insulator Bi$_{1.5}$Sb$_{0.5}$Te$_{1.7}$Se$_{1.3}$} 

\author{Sunghun Kim}
\email{kimsh@issp.u-tokyo.ac.jp}
\affiliation{Institute for Solid State Physics, The University of Tokyo, Kashiwa, Chiba 277-8581, Japan}
\author{Shunsuke Yoshizawa}
\affiliation{Institute for Solid State Physics, The University of Tokyo, Kashiwa, Chiba 277-8581, Japan}
\author{Yukiaki Ishida}
\affiliation{Institute for Solid State Physics, The University of Tokyo, Kashiwa, Chiba 277-8581, Japan}
\author{Kazuma Eto}
\affiliation{Institute of Scientific and Industrial Research, Osaka University, Ibaraki, Osaka 567-0047, Japan}
\author{Kouji Segawa}
\affiliation{Institute of Scientific and Industrial Research, Osaka University, Ibaraki, Osaka 567-0047, Japan}
\author{Yoichi Ando}
\email{y_ando@sanken.osaka-u.ac.jp}
\affiliation{Institute of Scientific and Industrial Research, Osaka University, Ibaraki, Osaka 567-0047, Japan}
\author{Shik Shin}
\affiliation{Institute for Solid State Physics, The University of Tokyo, Kashiwa, Chiba 277-8581, Japan}
\author{Fumio Komori}
\email{komori@issp.u-tokyo.ac.jp}
\affiliation{Institute for Solid State Physics, The University of Tokyo, Kashiwa, Chiba 277-8581, Japan}

\date{October 11, 2013}
\begin{abstract}

Electron scattering in the topological surface state (TSS) of the
bulk-insulating topological insulator
Bi$_{1.5}$Sb$_{0.5}$Te$_{1.7}$Se$_{1.3}$ was studied using quasiparticle
interference observed by scanning tunneling microscopy. It was found
that not only the 180$^{\circ}$ backscattering but also a wide range of
backscattering angles of 100$^{\circ}$--180$^{\circ}$ is effectively
prohibited in the TSS. This conclusion was obtained by comparing the
observed scattering vectors with the diameters of the constant-energy
contours of the TSS, which were measured for both occupied and
unoccupied states using time- and angle-resolved photoemission
spectroscopy. The unexpectedly robust protection from backscattering in
the TSS is a good news for applications, but it poses a challenge to the
theoretical understanding of the transport in the TSS.

\end{abstract}



\maketitle 

Three-dimensional (3D) topological insulators (TIs) are accompanied by
gapless surface states due to a nontrivial $Z_2$ topology of the bulk
wave functions \cite{Hasan,Qi,Ando}. Such a topological surface state
(TSS) of a 3D TI is peculiar in that it is helically spin polarized,
which leads to a suppression of electron scatterings due to spin
mismatch between the eigenstates before and after the scattering
\cite{Hasan,Ando}; in particular, 180$^\circ$ backscattering is expected
to be absent, because $+\mathbf{k}$ and $-\mathbf{k}$ states have
completely opposite spins. Such a protection of the TSS from
backscattering has been elucidated to play a key role in maintaining a
high mobility of the carriers in the TSS \cite{Taskin12}, and this
characteristic is one of the reasons why TIs are an appealing platform
for various device applications \cite{Moore}.

In scanning tunneling microscopy (STM) experiments, the suppression of
backscattering in TIs has been inferred from the measurements of 
quasiparticle interference (QPI) in Bi$_{1-x}$Sb$_{x}$ \cite{Roushan},
Bi$_{2}$Te$_{3} $ \cite{Zhang,Alpichshev}, and Bi$_{2}$Se$_{3}$
\cite{Hanaguri}, as well as in Ca- and Mn-doped Bi$_{2}$Te$_{3}$ and
Bi$_{2}$Se$_{3}$ \cite{Beidenkopf}. In Bi$_{1-x}$Sb$_{x}$ which has
multiple surface bands, strong interference was observed only between
those surface bands that have the same spin orientation. On the other
hand, in Bi$_{2}$Te$_{3}$ and Bi$_{2}$Se$_{3}$ where the TSS consists of
a single Dirac cone, no interference has been detected unless the Dirac
cone becomes significantly warped at energies away from the Dirac point
\cite{Zhang,Alpichshev,Hanaguri,Beidenkopf,Zhou}; remember, in those
materials a term proportional to $k^3 \sigma_z$ describing the cubic
spin-orbit coupling shows up in the effective Hamiltonian \cite{Fu},
which results in hexagonal warping of the Dirac cone that has been
experimentally observed by angle-resolved photoemission spectroscopy
(ARPES) \cite{Chen,Kuroda}.

In those previous experiments, the suppression of backscattering due to
the helical spin polarization of the TSS was only qualitatively
elucidated, because no QPI was observed for intraband scatterings in
Bi$_{1-x}$Sb$_{x}$ nor for circular constant-energy contours of the TSS
in Bi$_{2}$Te$_{3}$ and Bi$_{2}$Se$_{3}$. In other words, it has not
been clear to what extent the backscattering is suppressed as a function
of scattering angle when the scattering takes place within the same
surface band whose constant-energy contour is not warped. Even though such
information is crucial for establishing concrete understanding of the
transport in the TSS, until recently no TI material has allowed us to
observe QPI in an unwarped Dirac cone.

This situation has changed with the recent discovery of the
bulk-insulating TI material Bi$_{2-x}$Sb$_{x}$Te$_{3-y}$Se$_{y}$
\cite{Ren,Taskin11}. This is the first TI material in which the
surface-dominated transport was achieved in bulk single crystals
\cite{Taskin11}, and the Fermi energy $E_{\rm F}$ can be tuned in the
bulk band gap by changing the composition along particular combinations
of $(x,y)$ to realize the bulk-insulating state \cite{Ren,Arakane}. The
alloyed nature of this material is expected to cause long-range
potential fluctuations \cite{Roushan}, which would lead to relatively
strong scattering of long-wave-length electrons (i.e. Bloch electrons
with small $k$), even though the surface carrier mobility is high enough
to present clear Shubnikov-de Haas oscillations \cite{Taskin11}. Indeed, a
recent STM study of Bi$_{1.5}$Sb$_{0.5}$Te$_{1.7}$Se$_{1.3}$ found QPI
to be observable \cite{Ko} despite the weak warping of the Dirac cone. Therefore,
Bi$_{2-x}$Sb$_{x}$Te$_{3-y}$Se$_{y}$ offers a promising platform for
quantitatively understanding one of the most important characteristics of
TIs, the suppression of backscattering, in a nearly ideal Dirac cone. 

In this Letter, we elucidate how the elastic scattering among the
helically-spin-polarized surface electrons of
Bi$_{1.5}$Sb$_{0.5}$Te$_{1.7}$Se$_{1.3}$ is suppressed as a function of
the scattering angle and electron energy in the unwarped portion of the
Dirac cone. Such information became available because elastic scattering
of electrons manifests itself in the QPI down to energies close to the
Dirac-point energy $E_{\rm D}$, thanks to the long-range potential
fluctuations in this material. We found that there is a sharp threshold
for the length of the scattering vector, above which the QPI intensity
is abruptly diminished. Such a threshold points to the existence of a
well-defined critical scattering angle beyond which elastic scattering
is suddenly suppressed. By comparing the length of the critical
scattering vector in the QPI with the diameters of the constant-energy
contours of the TSS, we found that the maximum scattering angle is
$\sim$100$^\circ$ and is independent of the energy location, as long as
the bulk scattering channel does not intervene. For this comparison, we
measured the dispersions of the unoccupied states by using time-resolved
ARPES (TrARPES) implementing a pump-probe method, because in this
material $E_{\rm F}$ is located very close to $E_{\rm D}$ and a major
part of the upper Dirac cone is unmeasurable with usual ARPES. 

The single crystals of Bi$_{1.5}$Sb$_{0.5}$Te$_{1.7}$Se$_{1.3}$ were
grown by melting stoichiometric mixtures of high-purity elements in
sealed quartz tubes as described in Ref. \cite{Ren}. The crystal structure
was confirmed by x-ray diffraction. Experiments using STM and TrARPES
were performed in two separated ultra-high-vacuum (UHV) systems. After a
clean surface was prepared by cleaving the crystal in UHV of better than
2 $\times$ 10$^{-8}$ Pa at room temperature, the sample was
transferred {\it in situ} in UHV either to the cooled stage in the STM
chamber or to the TrARPES chamber.

The STM images and tunneling spectra were taken at 5 K using a cryogenic STM with an
electrochemically-etched W tip. The tip apex and its metallic density of
states were checked by scanning a clean Pt(111) surface. Topographic
images were obtained using a constant-current mode. For the study of
QPI, differential-conductance ($dI/dV$) spectroscopy was performed using
a standard lock-in technique with a bias-voltage modulation of 5--10
mV$_{\rm rms}$ at 496 Hz. The $dI/dV$ curve was measured at every points
of a 256 $\times$ 256 grid on the surface. The obtained data were
plotted as a function of position to make a $dI/dV$ map. The
constant-energy $dI/dV$ maps were Fourier transformed (FT), and the
peaks due to the surface lattice were used for the calibration of the
wave-vector space. The FT patterns were symmetrized with respect to the 6-fold symmetry, on the basis of the $C_{3v}$ symmetry of the cleaved surface.

In the TrARPES experiments, we used 1.5 eV (pump) and 5.9 eV (probe)
pulsed photons from an amplified Ti:Sapphire laser system with
repetition rate of 250 kHz \cite{Ishida}. The pulse widths were 170 and
250 fs and the spot diameters were $\sim$0.4 and $\sim$0.2 mm for the
1.5 and 5.9 eV photons, respectively. The time delay between the pump and the probe pulses
 was optimized to clarify the unoccupied TSS
band in the bulk band gap \cite{Sobota}. Photoelectrons from the cleaved surface cooled to 5 K were
detected by a hemispherical analyzer. The energy resolution of the
photoelectrons was 15 meV.

\begin{figure}
\includegraphics[width=8cm]{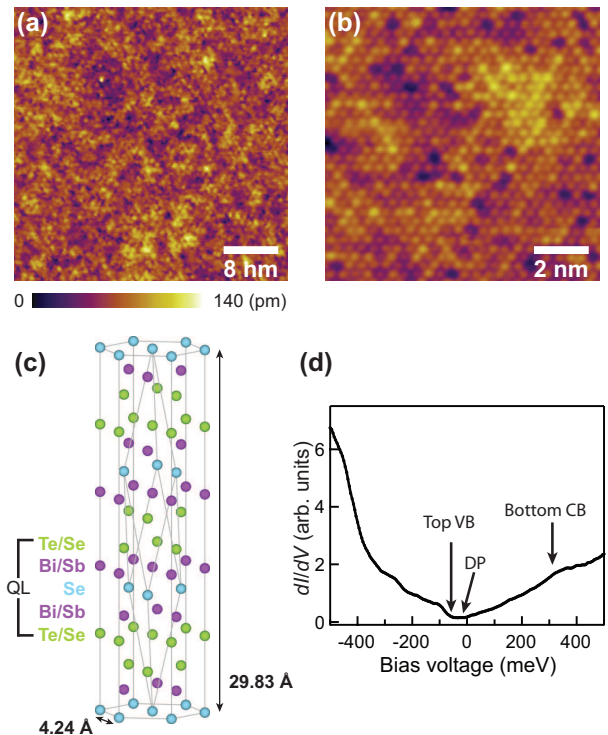}
\label{fig:STMSTS }
\caption{
(Color online) 
(a,b) STM images of a cleaved surface of
Bi$_{1.5}$Sb$_{0.5}$Te$_{1.7}$Se$_{1.3}$. Protrusions are Te or Se atoms
at the surface. The density of surface point defects is less than 3\%.
The sample bias voltage and tunneling current were $-0.4$ V and 20 pA in
(a), and $-0.6$ V and 100 pA in (b). (c) Crystal structure of
Bi$_{1.5}$Sb$_{0.5}$Te$_{1.7}$Se$_{1.3}$. 
(d) Typical differential conductance curve measured on a
cleaved surface at 5 K. The positions of the top of the bulk valence
band, the Dirac point, and the bottom of the bulk conduction band are marked
by arrows. The last is known from the result of TrARPES. 
}
\end{figure}

Typical STM images of the cleaved surface at 5 K are shown in Figs. 1(a)
and 1(b). In the magnified image, surface atoms are recognized with a
distribution of the apparent height. The
Bi$_{1.5}$Sb$_{0.5}$Te$_{1.7}$Se$_{1.3}$ crystal consists of
quintuple-layer units of (Te/Se)-(Bi/Sb)-Se-(Bi/Sb)-(Te/Se) that are
stacked and weakly bonded by the van-der-Waals force, as schematically shown in Fig.
1(c). The cleavage occurs along the van-der-Waals gap, and the flat
cleaved surface is always the Te/Se layer. The local density of states
fluctuates because of the inhomogeneous distributions of Te and Se, as
well as those of Bi and Sb, in the crystal. Thus, the observed
distribution of the apparent height of the surface atoms is attributed
to the electronic effect due to the alloying in the Bi/Sb and Te/Se
layers.

An example of the point tunneling spectrum is shown in Fig. 1(d). The
Dirac-point energy $E_{\rm D}$ can be defined as the minimum in $dI/dV$ 
and varies over the surface (the variation
of the point spectra and the distribution of $E_{\rm D}$ are shown in
\cite{Supple}). The average location of $E_{\rm D}$ is 10 $\pm$ 15 meV
below $E_{\rm F}$. In the point tunneling spectra (Fig. 1(d) and 
\cite{Supple}), the differential conductance rapidly increases below
$E_{\rm D}$ compared to that above $E_{\rm D}$. This is because the top
of the bulk valence band, which can be recognized as a clear shoulder in
the $dI/dV$ curve in Fig. 1(d), is located just below $E_{\rm D}$. The
bottom of the bulk conduction band is, however, not very clear in the
tunneling spectrum. 

\begin{figure}
\includegraphics[width=8.5cm]{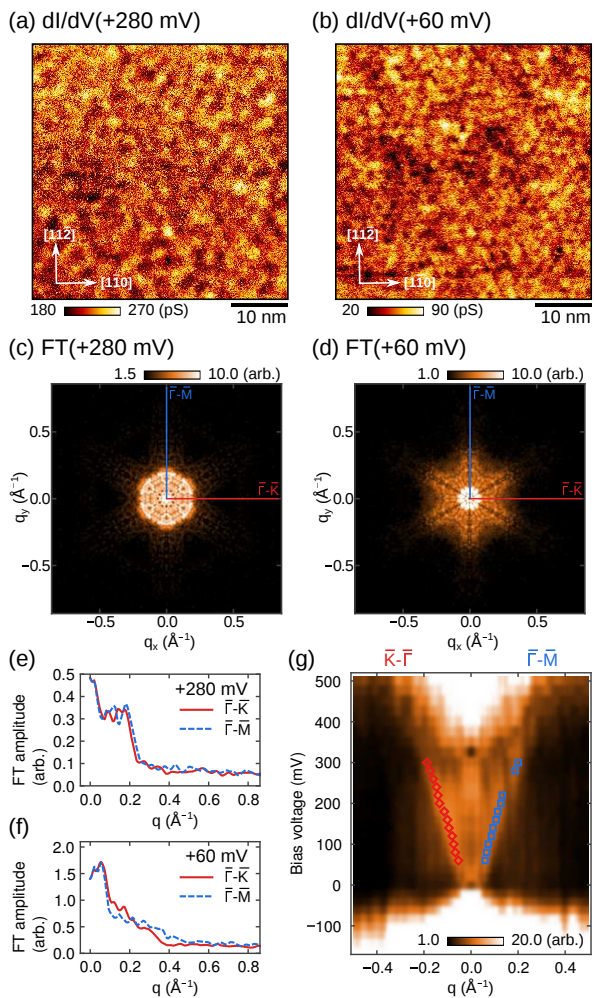}
\label{fig.QPI}
\caption{
(Color online) 
(a,b) Differential conductance images of the cleaved surface for $V_{\rm B}$
of (a) 280 mV and (b) 60 mV in the bulk band gap. (c,d) Corresponding FT
amplitude images. (e,f) Cross sections of the FT amplitude for the images
(c) and (d), respectively. The results in the $\bar{\Gamma}$-$\bar{\rm K}$
(solid, red) and $\bar{\Gamma}$-$\bar{\rm M}$ (dotted, blue) directions are
shown. (g) FT amplitude images as functions of $V_{\rm B}$ and the
scattering-vector length $q$ in the $\bar{\Gamma}$-$\bar{\rm K}$ (left) and
$\bar{\Gamma}$-$\bar{\rm M}$ (right) directions. Diamond (red) and square
(blue) symbols indicate the critical scattering-vector lengths in the two
directions. 
}
\end{figure}

Figures 2(a) and 2(b) show $dI/dV$ maps of the cleaved surface for two
selected sample bias voltages $V_{\rm B}$ in the bulk band gap, and the
corresponding FT images as shown in Figs. 2(c) and 2(d) (additional
images for various energies are shown in \cite{Supple}). The
cross sections of the FT images at the two $V_{\rm B}$ values are show in
Figs. 2(e) and 2(f) for the two high-symmetry directions,
$\bar{\Gamma}$-$\bar{\rm K}$ and $\bar{\Gamma}$-$\bar{\rm M}$. In each
cross section, one notices a steep decrease of the scattering amplitude
with increasing scattering-vector length $q$. Such a steep decrease in
the scattering amplitude was observed in the $V_{\rm B}$ range of 60--300
mV. This range corresponds to the energy window where the TSS is located
in the bulk band gap. Our data indicate that the electron scattering
within the TSS is rapidly diminished when the scattering-vector length
exceeds a certain critical value. Hence, we call it critical
scattering-vector length and denote it $q_{\rm cx}$ and $q_{\rm cy}$ for
$\bar{\Gamma}$-$\bar{\rm K}$ and $\bar{\Gamma}$-$\bar{\rm M}$ directions,
respectively. Obviously, they both increase with increasing $V_{\rm B}$; to
substantiate this trend, the FT amplitudes in the
$\bar{\Gamma}$-$\bar{\rm K}$ and $\bar{\Gamma}$-$\bar{\rm M}$ directions are
shown in Fig. 2(g) as an image on the scattering-vector length $q$ vs
$V_{\rm B}$ plane, and the critical scattering-vector lengths are marked by
red and blue symbols on the image. 

\begin{figure}
\includegraphics[width=6.5cm]{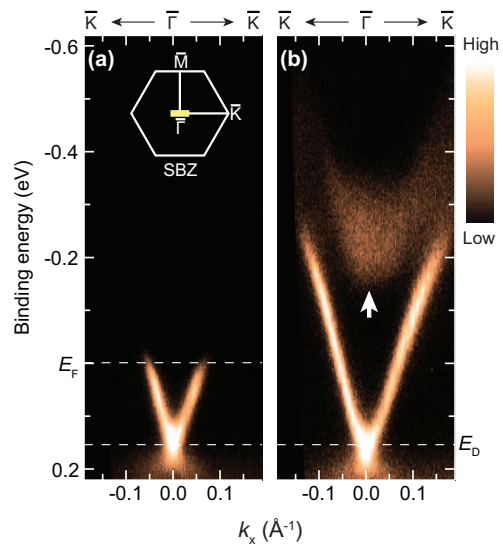}
\label{fig.ARPES}
\caption{
(Color online) 
(a,b) Photoemission intensity images of the cleaved surface of
Bi$_{1.5}$Sb$_{0.5}$Te$_{1.7}$Se$_{1.3}$ plotted along the
$\bar{\Gamma}$-$\bar{\rm K}$ direction; images (a) and (b) were obtained
without and with the 1.5-eV pump photons, respectively. 
Dispersions above the Fermi level became visible by pumping electrons 
into the unoccupied side. The bottom of the unoccupied bulk conduction band is
indicated by an arrow. Inset of (a) shows the surface Brillouin zone,
together with the range covered by the TrARPES measurements shown as a
short thick line. 
}
\end{figure}

To quantitatively understand the implications of the observed critical
scattering-vector lengths, information of the TSS dispersion above
$E_{\rm F}$ is important. Such information is unavailable with the
ordinary ARPES, but the TrARPES makes it possible to measure it with a
high resolution. To demonstrate the power of TrARPES, Figs. 3(a) and
3(b) compares the results of ARPES and TrARPES in the
$\bar{\Gamma}$-$\bar{\rm K}$ direction at 5 K. The band dispersion of the
TSS was observed up to 0.25 eV above $E_{\rm F}$ as shown in Fig. 3(b);
the unoccupied states at the center of the surface Brillouin zone in
this figure are naturally assigned to the bulk states, on the basis of
the electronic states of Bi$_{2}$Te$_{3}$ and Bi$_{2}$Se$_{3}$
\cite{Chen,Kuroda}. The band shape of the TSS shown in Fig. 3(a) is
consistent with the previous result \cite{Arakane}. We note that $E_{\rm
D}$ in those ARPES data is located at 0.15 eV below $E_{\rm F}$, which
is lower than that observed in the tunneling spectrum shown in Fig.
1(d). This difference can be explained by the electron doping from
adsorbed hydrogen that often occurs during ARPES measurements at low
temperature \cite{Kaminski}; it has been elucidated that residual
hydrogen molecules are dissociatively adsorbed on the surface by
irradiation of ultra-violet light. 

The velocity of the surface band is 3.5 $\times$ 10$^5$ m/s just above
$E_{\rm D}$, which increases to 4.9 $\times$ 10$^5$ m/s at 80 meV above
$E_{\rm D}$. Such a change in the surface band velocity also occurs in
Bi$_2$Se$_3$ \cite{Ando,Chen2,Kuroda2}. No significant difference in the band
dispersion was observed between the $\bar{\Gamma}$-$\bar{\rm M}$ and
$\bar{\Gamma}$-$\bar{\rm K}$ directions by TrARPES for energies up to 100
meV above $E_{\rm F}$ within our experimental accuracy. This means that
the cross section of the TSS is close to circular in this energy range,
and warping would become noticeable only at higher energies.

\begin{figure}
\includegraphics[width=7.0cm]{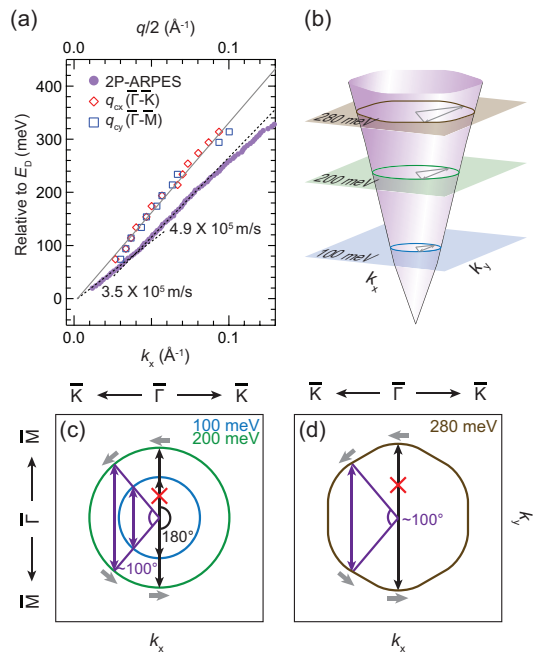}
\label{fig.analysis}
\caption{
(Color online)  
(a) The critical scattering-vector lengths $q_{\rm cx}$ and $q_{\rm cy}$
and the diameter of constant-energy contour of the TSS are plotted for
varying energy from $E_{\rm D}$ to $E _{\rm D} + 310$ meV; $q_{\rm cx}$
and $q_{\rm cy}$ were obtained from Fig. 2(g), and the diameter of the
TSS in the $\bar{\Gamma}$-$\bar{\rm K}$ direction was calculated from Fig.
3(b). (b) Schematic picture of the energy-dependent shape of the upper
Dirac cone together with the available scattering vectors at
representative energies. (c,d) Schematic pictures for circular and
warped TSS, respectively, to indicate that the critical
scattering-vector lengths being 75\% of the diameter of TSS corresponds
to the maximum scattering angle of 100$^{\circ}$; any scattering with a
larger angle requires longer scattering vectors and hence is prohibited.
}
\end{figure}

Knowing the TSS dispersion above $E_{\rm F}$ for
Bi$_{1.5}$Sb$_{0.5}$Te$_{1.7}$Se$_{1.3}$, we are now in the position to
make quantitative analysis of the critical scattering-vector lengths
found in the QPI data. In Fig. 4(a), $q_{\rm cx}$ and $q_{\rm cy}$ are
compared with the diameter of the cross section of the Dirac-cone TSS
observed by TrARPES. Here, the origin of the energy is commonly set to
be $E_{\rm D}$. The critical scattering-vector lengths is about 75\% of
the corresponding diameter of the TSS at any energy between 90 and 310
meV above $E_{\rm D}$. Note that if scattering is allowed for the
scattering angle of up to 180$^{\circ}$ (i.e. no restriction for
backscattering), $q_{\rm c}$ should be equal to the diameter of the TSS.
Hence, the fact that $q_{\rm c}$ is limited to 75\% of the diameter of
the TSS means that the allowed scattering angle has a maximum, which can
be easily calculated to be 100$^\circ$. This situation is schematically
depicted in Figs. 4(c) and 4(d) for circular and warped cross sections of
the TSS. Figure 4(b) graphically shows that this maximum scattering
angle does not change with energy. This result indicates that not only
the 180$^{\circ}$ backscattering but also a rather wide range of
backscattering angle of 100$^{\circ}$--180$^{\circ}$ is effectively
prohibited due to the spin mismatch between the initial and final states
in the TSS. This is a good news for applications to utilize the
protection of the TSS from backscattering. Naturally, theoretical
models to account for this unexpectedly robust protection from
backscattering is strongly called for.

In summary, we found critical scattering-vector lengths in the QPI,
beyond which elastic scattering of electrons in the TSS is significantly
suppressed. The comparison with the TSS dispersions for the unoccupied
states obtained from TrARPES allowed us to conclude that the protection
from backscattering in the TSS occurs not only for 180$^{\circ}$ but
also a rather wide range of angles of 100$^{\circ}$--180$^{\circ}$.
Also, such a wide angle range for the protection from backscattering is
found to be essentially independent of the energy until the Dirac cone
becomes warped and/or the bulk scattering events intervene. At energies
higher than 300 meV, we found hexagonal patterns in the FT-QPI images
that come from warping of the Dirac cone, and in this energy range the
critical scattering vector was not clearly observed, indicating a
different mechanism of the protection from backscattering in the warped
Dirac cone.

The authors thank Y. Ozawa and T. Otsu for their improvement in the TrARPES 
measurements. This work was partly supported by JSPS (KAKENHI Nos. 21244048 and
25220708), MEXT (Innovative Area ``Topological Quantum Phenomena"
KAKENHI), and AFOSR (AOARD 124038).


\end{document}